\documentclass[a4paper,11pt]{article}
\usepackage{pos}
\usepackage{enumerate}

\def\beq{\begin{equation}}
\def\eeq{\end{equation}}
\def\bea{\begin{eqnarray}}
\def\eea{\end{eqnarray}}
\def\nn{\nonumber}
\def\roughly#1{\mathrel{\raise.3ex\hbox
{$#1$\kern-.75em\lower1ex\hbox{$\sim$}}}}

\def\sla#1{\raise.15ex\hbox{$/$}\kern-.57em #1}

\def\bctaunu{b \to c \tau^- {\bar\nu}_\tau}
\def\bclnu{b \to c \ell^- {\bar\nu}_\ell}
\def\bcmunu{b \to c \mu^- {\bar\nu}_\mu}

\def \mA0T{\mathcal{A}_{0, T}}

\def \mA0T{A_{0, T}}

\title{${\bar B} \to D^* \ell^- {\bar\nu}_\ell$ Decays: \\
    Angular Distributions and New Physics}
\ShortTitle{${\bar B} \to D^* \ell^- {\bar\nu}_\ell$ Decays:
    Angular Distributions and New Physics}

\author*{David London}

\affiliation{Physique des Particules, Universit\'e de Montr\'eal,\\
  1375 Avenue Th\'er\`ese-Lavoie-Roux, Montr\'eal, QC, Canada  H2V 0B3}

\emailAdd{london@lps.umontreal.ca}

\abstract{ABSTRACT: At the present time, there are hints of new
  physics (NP) in several observables involving $\bclnu$ decays. In
  this talk, I describe measurable angular distributions for ${\bar B}
  \to D^* \mu^- {\bar\nu}_\mu$ and ${\bar B} \to D^* \tau^- (\to \pi^-
  \nu_\tau) {\bar\nu}_\tau$ decays, including the most general NP
  contributions. These angular distributions contain enough
  information to pin down the Lorentz structure of the NP, which will
  help to identify it. They also have the ability to reveal the
  presence of non-SMEFT (non-decoupling) NP.}

\FullConference{%
  11th International Workshop on the CKM Unitarity Triangle (CKM2021)\\
  22-26 November 2021\\
  The University of Melbourne, Australia
}

\begin{document}

\begin{flushright}
  {UdeM-GPP-TH-22-293}
\end{flushright}

\maketitle

{\bf\boldmath $\bctaunu$ Anomalies \cite{London:2021lfn} ---} Several
observables have been measured that involve $\bclnu$ decays:
\beq
R_{D^{(*)}} \equiv \frac{ {\cal B}({\bar B} \to D^{(*)} \tau^- {\bar\nu}_\tau) } { {\cal B}({\bar B} \to D^{(*)} \ell^- {\bar\nu}_\ell) } ~,~ \ell=e,\mu ~~,~~~~  
R_{J/\psi} \equiv \frac{ {\cal B}(B_c \to J/\psi \tau \nu_\tau)} { {\cal B}(B_c \to J/\psi \mu \nu_\mu)} ~.
\eeq
The latest results for $R_{D^{(*)}}$ can be found on the HFLAV site
\cite{HFLAV}; the combined $R_D$ and $R_{D^*}$ measurements differ
from the SM predictions by $3.1\sigma$. As for $R_{J/\psi}$, the
discrepancy with the SM is at the level of $2.6\sigma$
\cite{Watanabe:2017mip}. These $\bclnu$ measurements hint at
$\tau$-$\mu$ and $\tau$-$e$ universality violation. The simplest
explanation is that there is new physics (NP) in $\bctaunu$ decays.

With this in mind, there are two other relevant observables using $B
\to D^* \tau \nu_\tau$. They are the $\tau$ polarization asymmetry
$P_\tau(D^*)$ and the longitudinal $D^*$ polarization $F_L(D^*)$,
defined as
\bea
\label{bctaunu_observ2}
& P_\tau(D^*) \equiv \frac{ \displaystyle \Gamma( B \to D^* \tau^{\lambda=+1/2} \nu_\tau ) - \Gamma( B \to D^* \tau^{\lambda=-1/2} \nu_\tau )}
{\displaystyle \Gamma( B \to D^* \tau \nu_\tau ) } 
~, & \nn\\
& F_L(D^*) \equiv \frac{\displaystyle \Gamma( B \to D^*_L \tau \nu_\tau ) }{\displaystyle \Gamma( B \to D^* \tau \nu_\tau ) } ~. &
\eea
$P_\tau(D^*)$ and $F_L(D^*)$ were measured in
Refs.~\cite{Belle:2016dyj, Belle:2017ilt} and \cite{FLmeas},
respectively. A fit to all the data was performed in
Ref.~\cite{Blanke:2019qrx}. It was found that, for the SM, the
$p$-value of the fit is $\sim 0.1\%$, corresponding to a discrepancy
of $3.3\sigma$.

{\bf New-Physics Explanations ---} There are two different approaches
to examining possible NP explanations of these anomalies. The first is
to construct specific NP models, and here three classes of models have
been proposed \cite{London:2021lfn}, those involving (i) a new $W'$
boson, (ii) a leptoquark, or (iii) a charged Higgs boson.

The second is an effective field theory (EFT) approach. Assuming a
left-handed (LH) neutrino, the most general effective Hamiltonian for
$\bctaunu$ is
\bea
\label{Heff}
H_{\rm eff} &=& \frac{4 G_F}{\sqrt{2}} \, V_{cb} \left\{ 
\left[ (1 + C_V^L) ( {\bar c} \gamma^\mu P_L b ) + C_V^R ( {\bar c} \gamma^\mu P_R b ) \right] 
( {\bar \tau} \gamma_\mu P_L \nu_\tau ) \right. \\
&& \hskip2truecm \left.
+ \left[ C_S^R ( {\bar c} P_R b ) + C_S^L ( {\bar c} P_L b ) \right] ( {\bar \tau} P_L \nu_\tau ) 
+ C_T ( {\bar c} \sigma^{\mu\nu} P_L b ) ( {\bar \tau} \sigma_{\mu\nu} P_L \nu_\tau ) \right\} + h.c. \nn
\eea
Here, the coefficients $C_{V,S,T}^{L,R}$ are generated only through NP.

The question then is: how can we distinguish among these NP
contributions? In this talk, I suggest using angular distributions, as
these are sensitive to different Lorentz structures.

\begin{figure}[t]
\begin{center}
  \includegraphics[width=0.6\hsize]{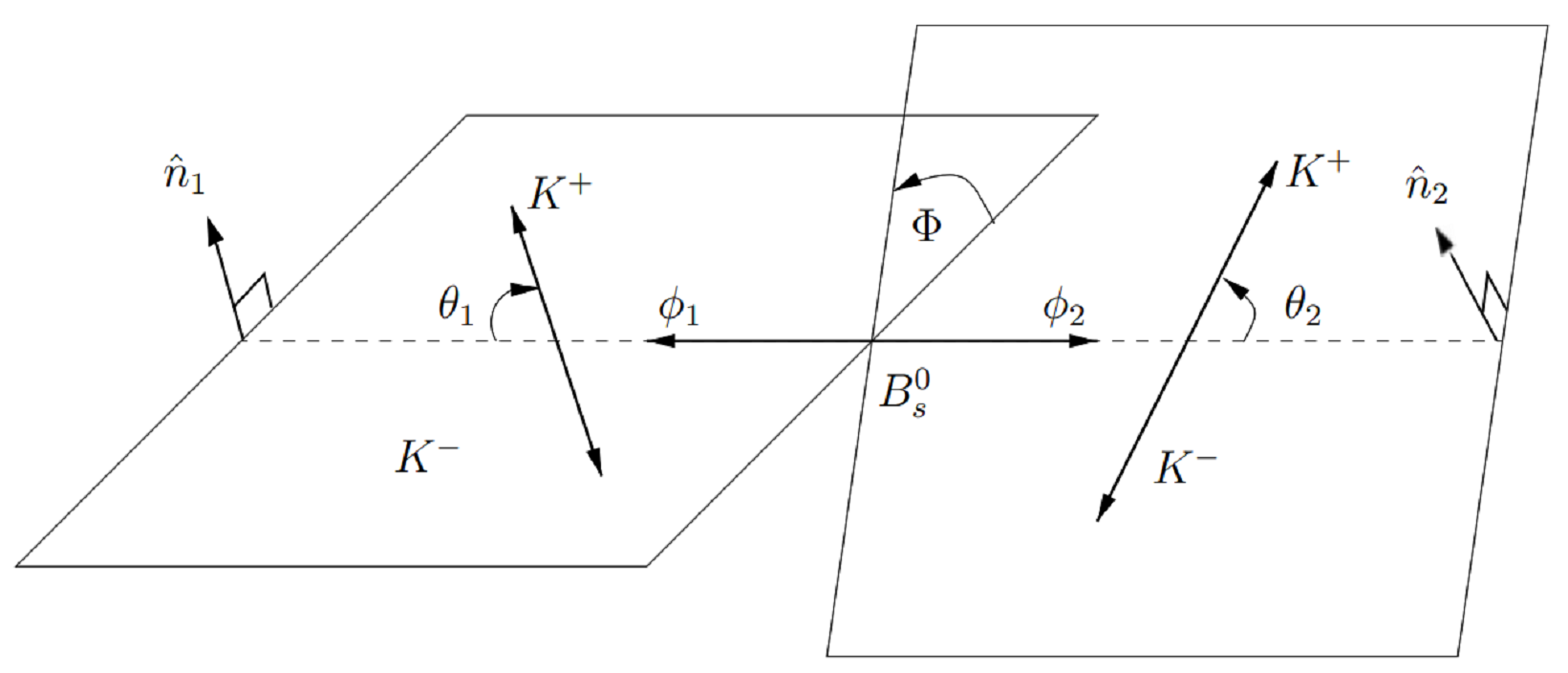}
\end{center}
\caption{Definition of the angles in the $B_s^0 \to \phi \phi$ angular
  distribution.}
\label{BVVangles}
\end{figure}

{\bf Angular Distributions ---} Consider $B$ decays of the form $B \to
V_1 (\to P_1 P'_1) V_2 (\to P_2 P'_2)$.  Examples of these include
$B_d \to \phi K^*$ and $B_s^0 \to \phi \phi$. In these decays, there
are three helicity amplitudes, $A_0$, $A_\parallel$, and
$A_\perp$. Their presence leads to a number of different angular
functions in $|A(B \to V_1 (\to P_1 P'_1) V_2 (\to P_2
P'_2))|^2$. Specifically, the differential decay rate is a function of
three angles, $\theta_1$, $\theta_2$, and $\Phi$, shown in
Fig.~\ref{BVVangles}. 

The angular distributions in these decays are used mainly to search
for CP-violating effects. Since such effects are predicted to be small
in the SM, the measurement of CP violation in any of these decays
would be a smoking-gun signal of NP.

CP violation arises from the interference of two amplitudes. Consider
two helicity amplitudes $A_\perp$ and $A_i$ ($i=0$ or $\parallel$).
These can be written $A_\perp = |A_\perp| e^{i\phi_\perp}
e^{i\delta_\perp}$ and $A_i = |A_i| e^{i\phi_i} e^{i\delta_i}$, where
the $\phi_{\perp,i}$ are weak (CP-odd) phases and the
$\delta_{\perp,i}$ are strong (CP-even) phases. Strong phases are
typically produced due to QCD effects, and since $B \to V_1 (\to P_1
P'_1) V_2 (\to P_2 P'_2)$ decays are purely hadronic, the helicity
amplitudes may well have different strong phases. 

When one computes $|A|^2$, the term $\epsilon_{\mu\nu\rho\sigma}
p_1^\mu p_2^\nu p_3^\rho p_4^\sigma$ arises.  In the rest frame of,
for example, particle \#1, this equals $m_1 {\vec p}_2 \cdot ({\vec
  p}_3 \times {\vec p}_4)$, which is a triple product (TP). Its
coefficients (for $i=0,\parallel$) are
\beq
{\rm Im}(A_\perp A_i^*) = |A_\perp||A_i| (
\underbrace{ \sin(\phi_\perp - \phi_i) \cos(\delta_\perp - \delta_i)}_{{\hbox{\scriptsize CP-odd}}} 
+ \underbrace{ \cos(\phi_\perp - \phi_i) \sin(\delta_\perp - \delta_i)}_{{\hbox{\scriptsize CP-even}}} ) ~.
\label{TPcoeff}
\eeq
These are also the coefficients of the P-odd terms in the angular
distribution, which are proportional to $\sin\Phi$ ($\Phi$ is the
angle between the $V_1 \to P_1 P'_1$ and $V_2 \to P_2 P'_2$ decay
planes, see Fig.~\ref{BVVangles}). This implies that the corresponding
terms in the angular distribution of the CP-conjugate decay have the
opposite sign.

Note that a nonzero TP is not necessarily a signal of CP violation. If
there is no CP violation in the decay ($\phi_{\perp,0,\|}=0$), there
can still be a nonzero TP if $\delta_\perp - \delta_i \ne 0$ (see the
CP-even term in Eq.~(\ref{TPcoeff}) above), in which case we expect
that TP(decay) $= -$TP (CP-conjugate decay). Thus, a signal of CP
violation would be TP(decay) $+$ TP (CP-conjugate decay) $\ne 0$. To
be specific, this isolates the CP-odd term above, and indicates that
there are NP contributions to the decay.

{\bf\boldmath ${\bar B} \to D^* \mu^- {\bar\nu}_\mu$ ---} In the SM,
this decay takes place via ${\bar B} \to D^* (\to D \pi) W^* (\to
\mu^- {\bar\nu}_\mu)$. This is similar to $B \to V_1 (\to P_1 P'_1)
V_2 (\to P_2 P'_2)$, but there are two important differences: (i)
${\bar B} \to D^* \mu^- {\bar\nu}_\mu$ is a semileptonic decay,
implying that the hadronic and leptonic currents factorize (unlike $B
\to V_1 (\to P_1 P'_1) V_2 (\to P_2 P'_2)$), and (ii) the $W^*$ is
virtual, leading to four helicity amplitudes: $A_0$, $A_\parallel$,
$A_\perp$, $A_t$ (timelike).
  
Ref.~\cite{Bhattacharya:2019olg} examines the effect of adding NP to
this decay. The analysis assumes a LH neutrino, but is otherwise
completely general. It uses an EFT approach: the decay is generically
described by ${\bar B} \to D^* (\to D \pi) N^* (\to \mu^-
{\bar\nu}_\mu)$, and the NP effects are included in an effective
Hamiltonian similar to Eq.~(\ref{Heff}):
%
\begin{center}
\begin{tabular}{c|c|c|c|c}
coupling & quarks & leptons & spin & type \\
\hline
$(1 + C_V^L)$ & ${\bar c} \gamma_\mu P_L b$ & ${\bar \mu} \gamma^\mu P_L \nu_\mu$ & vector & SM + NP \\
$C_V^R$ & ${\bar c} \gamma_\mu P_R b$ & ${\bar \mu} \gamma^\mu P_L \nu_\mu$ & & NP \\
$C_S^R$ & ${\bar c} P_R b$ & ${\bar \mu} P_L \nu_\mu$ & scalar & NP \\
$C_S^L$ & ${\bar c} P_L b$ & ${\bar \mu} P_L \nu_\mu$ & & NP \\
$C_T$ & ${\bar c} \sigma^{\mu\nu} P_L b$ & ${\bar \mu} \sigma_{\mu\nu} P_L \nu_\mu$ & tensor & NP \\ 
\hline
\end{tabular}
\end{center}

The addition of general NP leads to four new helicity amplitudes:
$A_{SP}$, $A_{0,T}$, $A_{\parallel ,T}$, $A_{\perp ,T}$.  The helicity
amplitudes depend on the NP parameters as follows:
%
\begin{center}
\begin{tabular}{|c|c|} \hline
Helicity Amplitude & Coupling \\
\hline
$A_0$, $A_\parallel$, $A_t$ & $1+C_V^L-C_V^R$ \\
$A_\perp$ & $1+C_V^L+C_V^R$ \\
$A_{SP}$ & $C_P \equiv \frac12(C_S^R - C_S^L)$ \\
$A_{0,T}$, $A_{\parallel ,T}$, $A_{\perp ,T}$ & $C_T$ \\
\hline
\end{tabular}
\end{center}

Now, $C_V^L$, $C_V^R$, $C_P$ and $C_T$ each have a magnitude and a
weak (CP-odd) phase. However -- and this is the key point -- because
${\bar B} \to D^*$ is the only hadronic transition in the decay, all
helicity amplitudes are expected to have the same strong (CP-even)
phase. Thus, this process involves only seven NP parameters: the four
magnitudes of $1 + C_V^L$, $C_V^R$, $C_P$ and $C_T$, and their three
relative weak phases.

\begin{figure}[t]
\begin{center}
  \includegraphics[width=0.5\hsize]{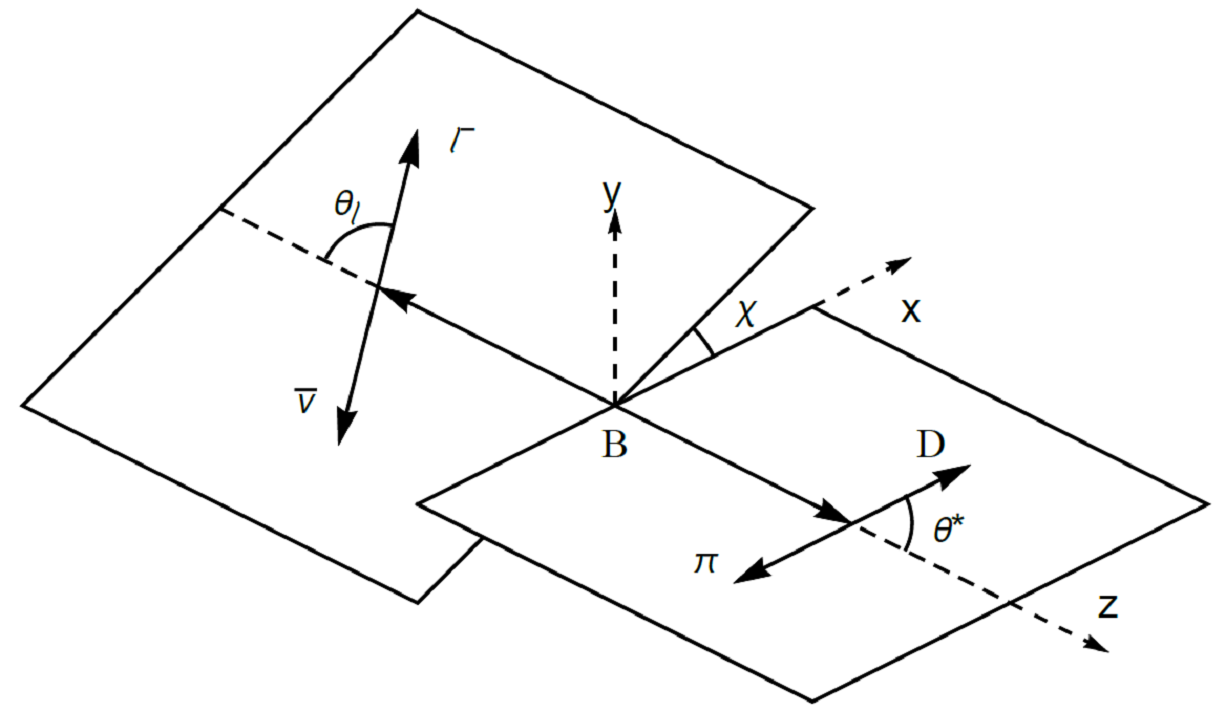}
\end{center}
\caption{Definition of the angles in the ${\bar B} \to D^* \mu^-
  {\bar\nu}_\mu$ angular distribution.}
\label{BD*munuangles}
\end{figure}

The differential decay rate can be computed as a function of the eight
helicity amplitudes. It is found that (i) the differential decay rate
is a function of $q^2$ and three angles, $\theta^*$, $\theta_\ell$,
$\chi$ (see Fig.~\ref{BD*munuangles}), and (ii) the angular
distribution contains twelve independent angular functions. Nine terms
are CP-conserving and are present in the SM. There are also three
CP-violating terms.  These are TPs, proportional to $\sin\chi$, and
arise only in the presence of NP. Furthermore, since the strong phases
of all helicity amplitudes are equal, the CP-even piece of
Eq.~(\ref{TPcoeff}) vanishes, so the measurement of a TP in ${\bar B}
\to D^* \mu^- {\bar\nu}_\mu$ is itself a signal of CP violation, i.e.,
of NP.
  
The bottom line is that, for each $q^2$ bin, there are twelve observables
(the coefficients of the angular functions). All of these are
functions of seven NP parameters.  With more observables than
unknowns, in principle, it is possible to extract {\it all} the
unknown parameters (this assumes that the value of $V_{cb}$ is known).

I want to emphasize the following: because $B \to V_1 (\to P_1 P'_1)
V_2 (\to P_2 P'_2)$ decays are purely hadronic, their angular
distribution can only be used to detect the presence of CP-violating
NP. On the other hand, since the ${\bar B} \to D^* \mu^-
{\bar\nu}_\mu$ decay is semileptonic, it is much cleaner. Here, the
angular distribution can also be used to detect the presence of
CP-conserving NP.

In Ref.~\cite{Belle:2017rcc}, Belle did an angular analysis of
${\bar B} \to D^* \ell^- {\bar\nu}_\ell$ with the purpose of
extracting $V_{cb}$. The fit did not allow for NP contributions. To be
specific, TP terms were not included. Our suggestion is to redo the
analysis, including all NP terms, both CP-conserving and
CP-violating. With a full fit, one might be able to measure or at
least constrain the seven $\bcmunu$ NP parameters.

Note: one might think that this is a lot of work for a decay process
that shows no hints of NP. However, in Ref.~\cite{Bobeth:2021lya},
the Belle ${\bar B} \to D^* \ell {\bar\nu}_\ell$ data was reanalyzed,
and it was found that there is some tension, possibly suggesting a
violation of $e$-$\mu$ universality. So perhaps it is worthwhile to
search for NP in this decay.

{\bf\boldmath ${\bar B} \to D^* \tau^- {\bar\nu}_\tau$ ---} The
measurement of the ${\bar B} \to D^* \mu^- {\bar\nu}_\mu$ angular
distribution requires the knowledge of $p_\mu$. For this reason, this
method cannot be applied to ${\bar B} \to D^* \tau^- {\bar\nu}_\tau$:
because the $\tau^-$ decays, $p_\tau$ is not measurable due to the
missing ${\bar\nu}_\tau$.  In Ref.~\cite{Bhattacharya:2020lfm}, an
alternative angular distribution is proposed that uses the decay
$\tau^- \to \pi^- \nu_\tau$. The process is then a series of two-body
decays: ${\bar B} \to D^* N^*$, $D^* \to D \pi'$. $N^* \to \tau^-
{\bar\nu}_\tau$, $\tau^- \to \pi^- \nu_\tau$.

\begin{figure}[t]
\begin{center}
  \includegraphics[width=0.5\hsize]{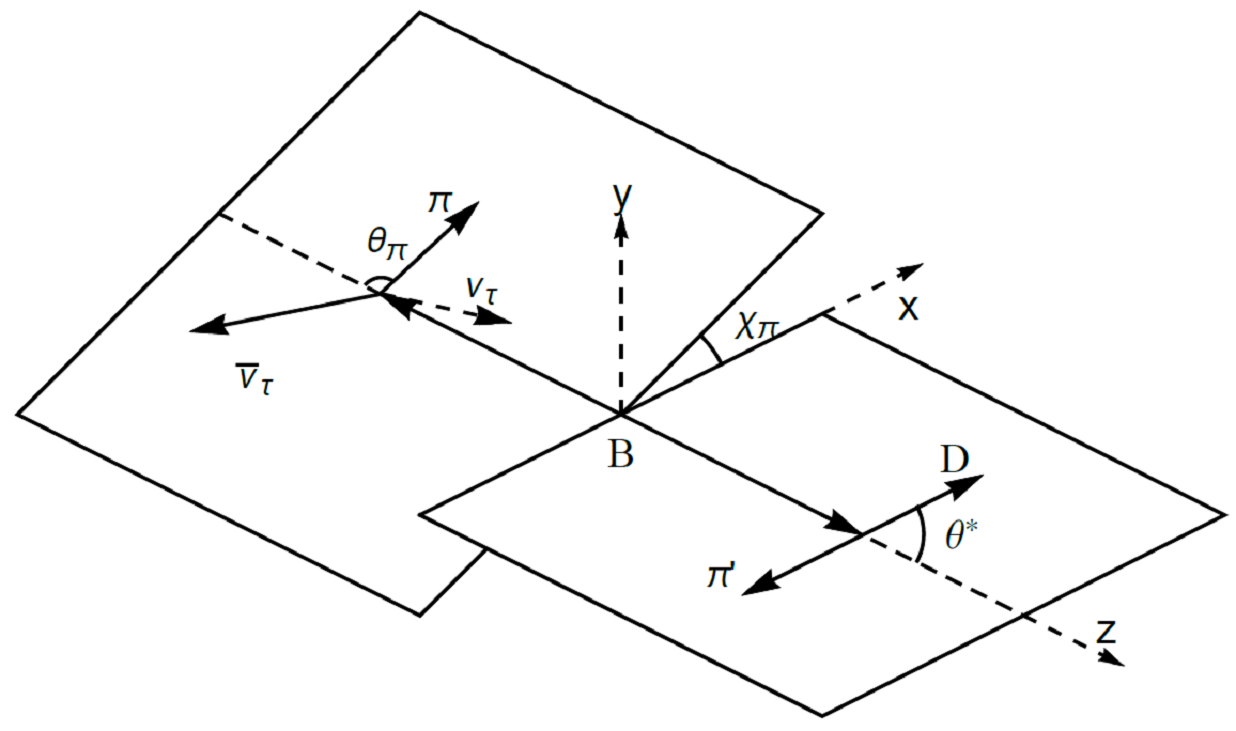}
\end{center}
\caption{Definition of the angles in the ${\bar B} \to D^* (\to D
  \pi') \, \tau^{-} (\to \pi^- \nu_\tau) {\bar\nu}_\tau$ angular
  distribution.}
\label{BD*tauunuangles}
\end{figure}

The $D^*$ and $\tau^-$ are on-shell, so that the final state is
described by six kinematic parameters. Due to the lack of knowledge of
$p_\tau$, not all of these are measurable. Even so, an angular
distribution can be constructed using the $\pi^-$ from the $\tau^-$
decay. $E_\pi$ and $q^2$ (the square of the momentum of the $\tau^-
{\bar\nu}_\tau$ pair) are measurable. Also measurable are
$\theta_\pi$, $\chi_\pi$ and $\theta^*$ (see
Fig.~\ref{BD*tauunuangles}). The remaining unmeasurable angle can be
integrated over.

The differential decay rate is then a function of $q^2$, $E_\pi$ and
three angles, $\theta^*$, $\theta_\pi$, $\chi_\pi$. The angular
distribution can be written as follows:
\beq
\sum^{9}_{i=1} f^R_i(q^2,E_\pi)\Omega^R_i(\theta^*,
\theta_\pi,\chi_\pi) + \sum^{3}_{i = 1}f^I_i(q^2,E_\pi)\Omega^I_i(\theta^*,
\theta_\pi,\chi_\pi) ~.
\eeq
The nine $f^R_i \Omega^R_i$ terms are CP-conserving and are present in
the SM. The $f^R_i$ contain $|A_i|^2$ and ${\rm Re}[A_i A_j^*]$
pieces. The three $f^I_i \Omega^I_i$ terms are CP-violating; the
$f^I_i$ contain ${\rm Im}[A_i A_j^*]$ pieces. These terms are TPs,
proportional to $\sin\chi_\pi$, and are smoking-gun signals of NP.

The point is that the seven $\bctaunu$ NP parameters can be
extracted. Experimentalists will of course determine which type of
analysis is best -- $q^2/E_\pi$ bins, a fit to all the data, etc. If
the discrepancy with the SM persists, this will tell us which NP
parameters are nonzero. In turn, the knowledge of the Lorentz
structure will be very helpful in identifying the NP.

\newpage

{\bf Non-SMEFT NP \cite{Burgess:2021ylu} ---} But there's more. 

Whatever it is, the NP is heavy, with a mass of $O({\rm TeV})$ or
higher. When it is integrated out, the EFT that is produced must
necessarily respect the $SU(2)_L \times U(1)_Y$ electroweak gauge
symmetry. This symmetry can be realized linearly, i.e., it is broken
via the Higgs mechanism, in which case we have SMEFT. Alternatively,
it can be realized nonlinearly, i.e., it is broken in another way, and
this corresponds to HEFT, for example.  Since the discovery of the
Higgs boson, SMEFT is the default assumption, but HEFT is still
possible.

It can only be determined experimentally whether a linear or nonlinear
realization provides a better description of Nature. These two options
can be distinguished through their predictions for the size of certain
low-energy dimension-six four-fermion operators.

In particular, consider the operator $C_V^R ( {\bar c} \gamma^\mu P_R
b ) ( {\bar \tau} \gamma_\mu P_L \nu_\tau )$ of Eq.~(\ref{Heff}).
HEFT predicts that $|C_V^R| \sim O(1)/\Lambda_{\rm NP}^2$, like the
other NP coefficients. On the other hand, SMEFT predicts $|C_V^R| \sim
O(1)/\Lambda_{\rm NP}^2 \times v^2/\Lambda_{\rm NP}^2$, where $v$ is
the Higgs vev. This is much smaller than the other NP coefficients. As
discussed above, it is possible to measure all NP parameters in the
angular distribution of ${\bar B} \to D^* (\to D \pi') \, \tau^{-}
(\to \pi^- \nu_\tau) {\bar\nu}_\tau$. If it is found that $|C_V^R|$ is
much larger than the SMEFT prediction, this will point to non-SMEFT
(non-decoupling) NP.

{\bf Conclusions ---} In summary, I have described measurable angular
distributions for ${\bar B} \to D^* \mu^- {\bar\nu}_\mu$ and ${\bar B}
\to D^* \tau^- (\to \pi^- \nu_\tau) {\bar\nu}_\tau$, including the
most general NP contributions. These semileptonic decays are described
by seven NP parameters.

Although unrelated by lepton flavour universality -- the $\tau$
decays, while the $\mu$ does not -- the two distributions have some
similarities: both contain twelve independent angular functions. Nine
of these are CP-conserving and are present in the SM. Three terms are
CP-violating (TPs) and are smoking-gun signals of NP. In both cases,
if the angular distributions can be measured, the seven NP parameters
can be extracted. In the case of ${\bar B} \to D^* \tau^- (\to \pi^-
\nu_\tau) {\bar\nu}_\tau$, this is particularly interesting, as it
would give us information about the NP hinted at in the $\bctaunu$
anomalies.

Finally, these angular distributions have the ability to reveal the
presence of non-SMEFT (non-decoupling) NP.

\medskip

{\bf Acknowledgments:} I thank B. Bhattacharya, A. Datta, S. Kamali,
C.P. Burgess, S. Hamoudou and J. Kumar for their collaboration on projects
discussed in this talk.  This work is supported in part by the Natural
Sciences and Engineering Research Council of Canada (NSERC).

\end{document}